\documentclass[a4paper,11pt]{article}

\usepackage{jinstpub} 
\usepackage{siunitx}
\usepackage{textcomp}
\usepackage{mathtools}
\usepackage{lineno}
\usepackage{dblfloatfix} 
\usepackage{soul}

\title{\boldmath The Apollo ATCA Design for the CMS Track Finder and the Pixel Readout at the HL-LHC}

\author[b]{A. Albert,}
\author[b]{Z. Demiragli,}
\author[b]{D. Gastler,}
\author[c]{K. Hahn,}
\author[b]{E. Hazen,}
\author[a]{P. Kotamnives,}
\author[c]{S. Noorudhin,}
\author[b]{A. Peck,}
\author[b]{J. Rohlf,}
\author[a]{C. Strohman,}
\author[a]{P. Wittich,}
\author[a,1]{R. Zou\note{speaker}}
\emailAdd{rz393@cornell.edu}

\affiliation[a]{Cornell University,\\Ithaca, NY, USA}
\affiliation[b]{Boston University,\\Boston, MA, USA}
\affiliation[c]{Northwestern University,\\Chicago, IL, USA}

\abstract{The challenging conditions of the High-Luminosity LHC  require tailored hardware designs for the trigger and data acquisition systems. The Apollo platform features a ``Service Module'' with a powerful system-on-module computer that provides standard ATCA communications and application-specific ``Command Module''s with large FPGAs and high speed optical fiber links. The CMS version of Apollo will be used for the track finder and the pixel readout. It features up to two large FPGAs and more than 100 optical links with speeds up to 25\,Gb/s. We study carefully the design and performance of the board by using customized firmware to test power consumption, heat dissipation, and optical link integrity. This paper presents the results of these performance tests, design updates, and future plans.}
 

\collaboration[c]{on behalf of the CMS Collaboration}

\proceeding{Topical Workshop on Electronics for Particle Physics TWEPP2021 \\
  September 20-24, 2021\\
  Online}

\begin{document}
\maketitle
\flushbottom
\vspace{-10pt}

\section{Introduction}
\label{sec:intro}

High-performance Advanced Telecommunications Computing Architecture (ATCA) blades are gaining popularity in high-energy physics experiments in view of the High-Luminosity LHC. The development for high-energy physics applications has proven to be challenging with many problems that need to be addressed (power, cooling, optical fiber management, communication interfaces, etc.). The Apollo platform aims to provide a simple hardware environment and firmware and software toolkit which can be used for the development of ATCA blades~\cite{Apollo2019}. It consists of a common ``Service Module" (SM) that handles standard ATCA communications as well as clock and power delivery, and an application-specific ``Command Module" (CM) with large FPGAs and many optical links to accommodate demanding algorithms and large data flow.

The Apollo SM is a standard-sized ATCA blade with a 7\,U$\times$180\,mm cutout to accommodate one full-sized or two half-sized CM boards. It features a Xilinx Zynq System-on-Module with embedded Linux OS for control, monitoring, and local DAQ functions, a CERN Intelligent Platform Management Controller (IPMC)~\cite{CERNIPMC} or OpenIPMC~\cite{OpenIPMC}, and a Wisconsin Ethernet Switch Module~\cite{ESM}. Standard commercial power entry and conditioning modules are used to deliver 12\,V$_\text{DC}$ at up to 30\,A to the CM. 

\section{Overview and CMS Applications}
The CMS experiment plans to have major detector upgrades for the High-Luminosity LHC~\cite{CMSDetector}, including a new silicon tracker~\cite{CMSTracker}. The Apollo platform will have two applications in the tracker back-end processing system: the Level-1 Track Finding (TF) system and the Inner Tracker Data Trigger and Control system (IT-DTC). 

The Level-1 TF reconstructs tracks from the Outer Tracker and calculates track parameters for the Level-1 trigger system. The algorithm requires substantial FPGA resources and is thus speculated to have a high power consumption. It is expected to require one or two Virtex Ultrascale+ FPGAs and 62 optical links at 25\,Gb/s. A total of 162 Apollo blades are expected to be used for the TF. 


The IT-DTC reads hits from the front-end ASIC chips and converts the data into a compact format. It also converts the trigger signal from the CMS trigger to tokens and transmits them back to the front-end. It is expected to require 72 optical links at 10\,Gb/s and 16 optical links at 25\,Gb/s. A total of 28 Apollo boards are expected to be used for the IT-DTC. 

To satisfy the requirements of the two applications, the CMS design of the Apollo CM can accommodate one or two large Xilinx Ultrascale+ FPGAs with 4-channel bi-directional and 12-channel uni-directional Firefly optical engines that sum up to more than 100 optical links with speed up to 25\,Gb/s. Two revisions of the design are planned before the final production. The CMS Apollo revision~1 was produced by the end of 2019. The revision~2 design is currently being produced and evaluated. This paper describes the evaluation process, shares what we learned, and presents the final performance of the revision~1 design. It also describes major updates of the revision~2 design and preliminary results on link integrity.

\section{Revision 1 Performance Tests}

The CMS revision~1 Apollo blade, as shown in Fig.~\ref{fig:rev1}, has two large FPGAs (an XCKU15P (KU15P) in an A1760 package and an XCVU7P (VU7P) in a B2104 package) and sites for 4-channel and 12-channel Firefly optical transceivers with up to 124 optical links with speed up to 25\,Gb/s. This section summarizes the evaluation procedure and the final results. 

\subsection{Link Integrity}
\begin{figure*}[ht!bp]
\begin{minipage}{0.50\textwidth}
\centering
\includegraphics[width=1.\textwidth]{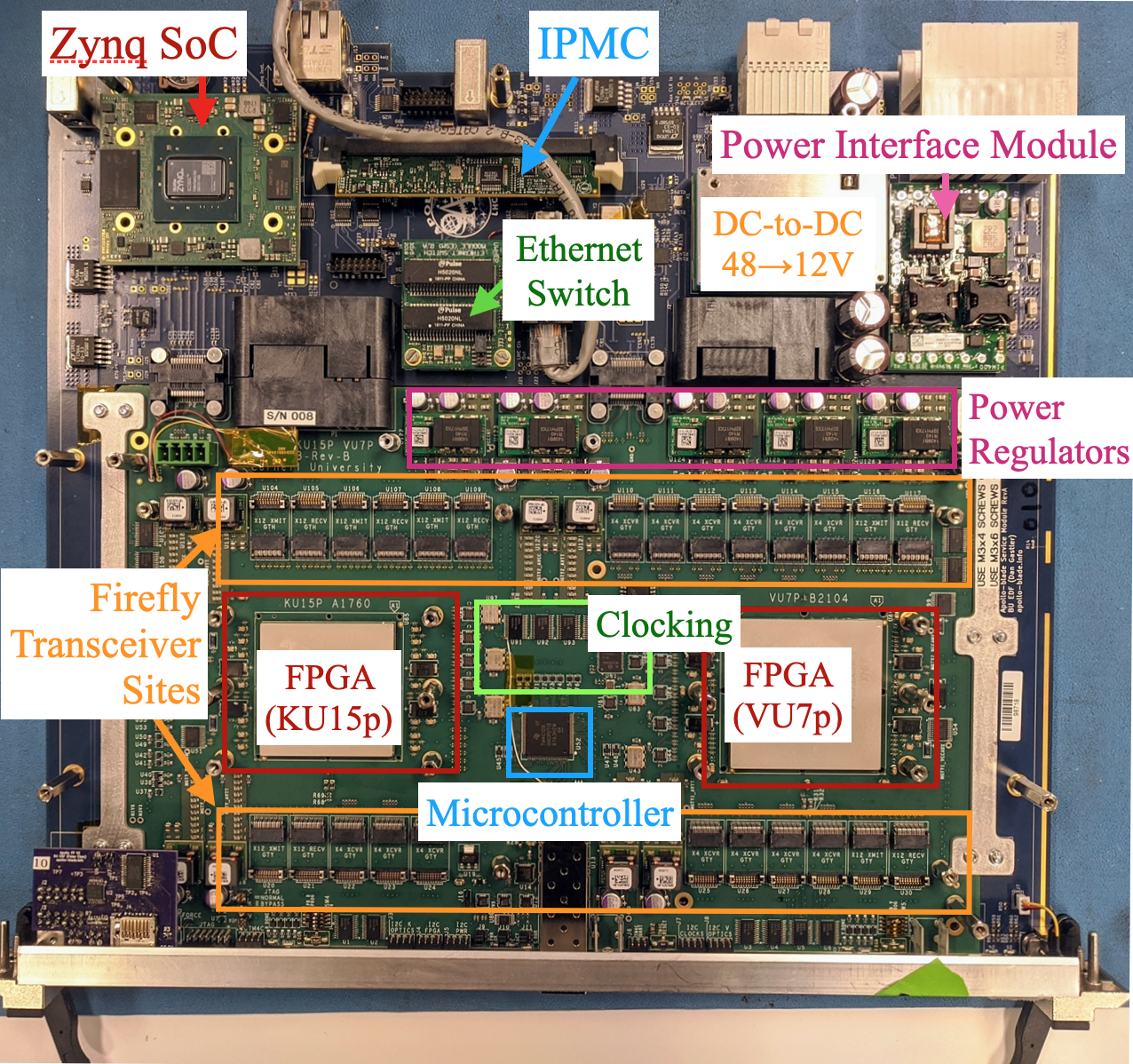}
\caption{\label{fig:rev1} The Apollo revision 1 SM and CM with major components labeled and highlighted in boxes.}
\end{minipage}
\qquad
\begin{minipage}{0.43\textwidth}
\centering
\includegraphics[width=1.\textwidth]{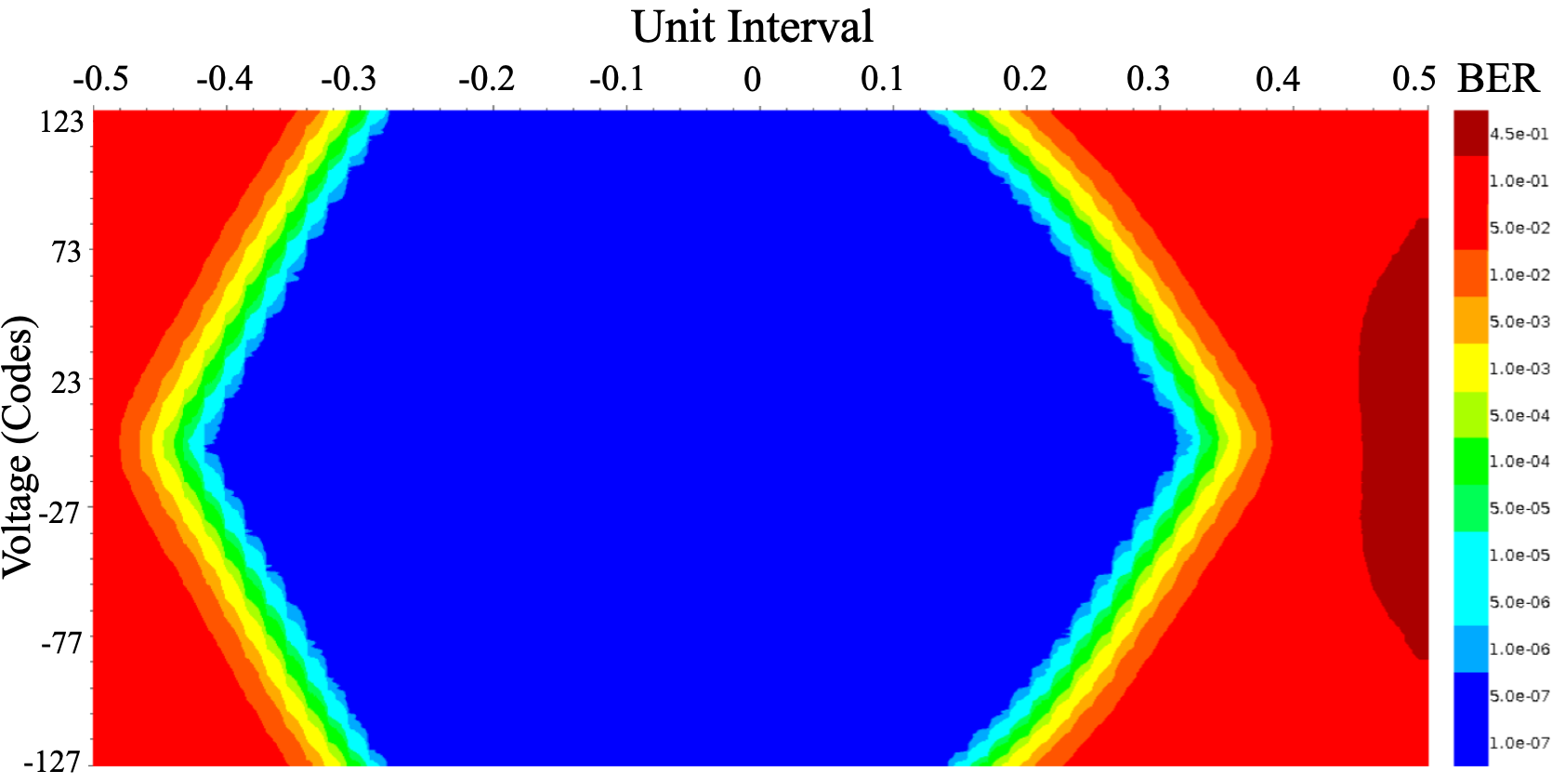}
\caption{\label{fig:linkintegrity} A typical eye pattern with a large opening at a data rate of 25\,Gb/s with CDR on the revision~1 CM.}
\end{minipage}
\end{figure*}

Up to 19 Firefly optical engines (13 4-channel parts and 6 12-channel parts) can be mounted on the revision~1 blade. We test the
entire data path by looping back the Multi-Gigabit Transceiver (MGT) on the board (MGT $\xrightarrow{\text{electrical}}$ Firefly Transmitter $\xrightarrow{\text{optical}}$ Firefly Receiver $\xrightarrow{\text{electrical}}$ MGT). The bit error rate  is measured to be less than $10^{-16}$ at 25\,Gb/s using the PRBS-31 pattern. Additional eye scans are conducted to check the receiver margins (Figure~\ref{fig:linkintegrity}). Because the clock data recovery (CDR) circuit is activated on the Firefly engines, the eye diagrams only give information on the electrical path between the Firefly receiver and the MGT.

\subsection{Power and Thermal Performance}
As the final algorithm firmware is under development, special power inefficient firmware, the Look-Up Table (LUT) oscillator algorithm~\cite{heaterpaper}, is used to test the power and thermal limits of the Apollo blade. The amount of power the FPGA exerts can be adjusted dynamically. Customizations per vendor's recommendations are applied to the LGA80D power supplies to achieve high output power to the FPGAs.  


We use the 450/40 FC Series 14-slot ATCA shelf which has a SCHROFF shelf manager with Pigeon Point ShMM 700. When the fan level goes from 10 to 15, shelf power consumption increases from 500\,W to 2000\,W, but the KU15P temperature only decreases from 77\,\textcelsius{} to 74\,\textcelsius{} at 80\,W. With the acoustic noise level also taken into consideration, we decided to set the maximum fan level to 10.

Since the VU7P heatsink is 20$\%$ bigger than the KU15P one, we expect the VU7P to have better thermal performance. However,  when the FPGAs were ﬁrst heated up, the VU7P was consistently warmer than the KU15P when drawing the same amount of power. Upon inspection, we found that the VU7P has a non-flat surface (0.1\,mm higher in the center), which makes good thermal conduct between the FPGA surface and the heatsink challenging.  A few thermal gels and putties are explored to gauge thermal conductivity and ease of application/removal. TG-PP-10 (10\,W/(mK)) is found to be the easiest to use and has the best conductivity. The VU7P is 8\,\textcelsius{} cooler at 70\,W with the TG-PP-10 than with the thermal pad (TG-A6200). 

Large temperature gradients across a single FPGA are observed in firmware designs with a concentrated number of LUT oscillators. The VU7P has two Super-Logic Regions (SLRs). Each SLR is a single FPGA die slice and has its own system monitor that can conduct an independent temperature measurement. With a concentrated number of LUT oscillators built on top of one of the two system monitors on the VU7P, a temperature difference of more than 40\,\textcelsius{} is observed between the two system monitors.  Further investigation suggests that large temperature gradients can also occur within a single SLR. Similar observations have been reported previously~\cite{heaterpaper}.

\begin{figure}[htbp]
\centering 
\resizebox{.9\textwidth}{!}{%
\includegraphics[height=3cm]{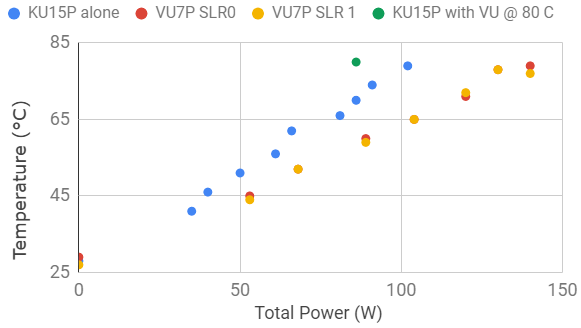}%
\quad
\includegraphics[height=3cm]{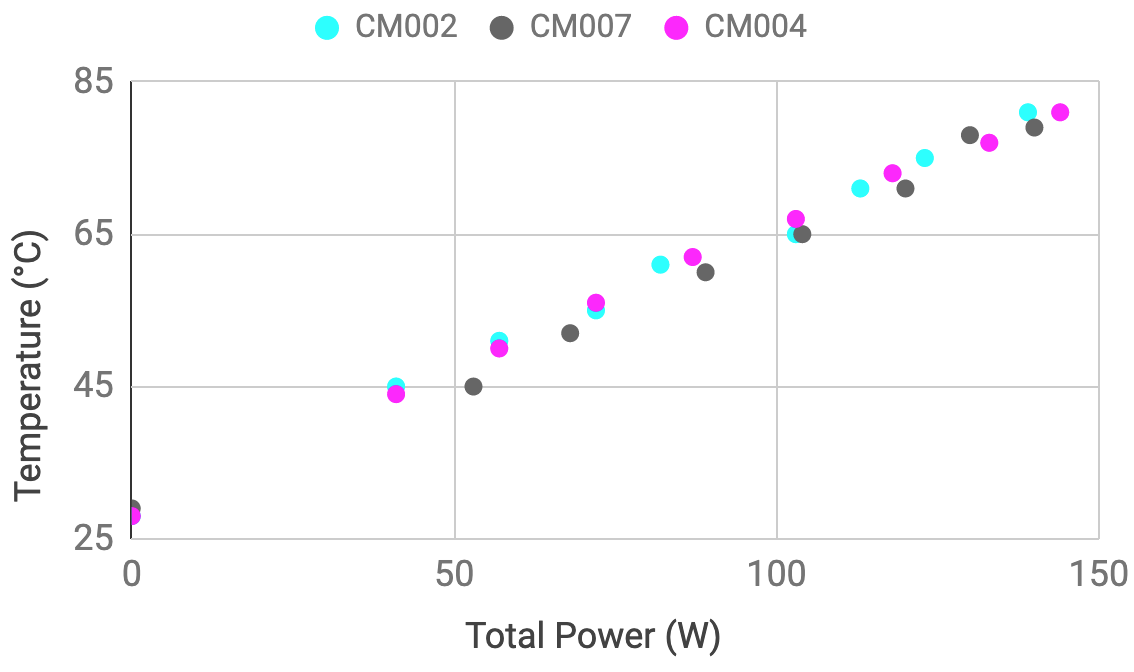}%
}
\vspace{-7pt}
\caption{\label{fig:finalperf} Temperature vs power performance of both FPGAs on one revision~1 CM (left) and the VU7Ps on three revision~1 CMs (right). Red and yellow dots are temperatures measured on the two SLRs of the VU7P. Blue dots represent the performance of the KU15P with the VU7P at 28\,\textcelsius{}. The green dot shows the reduced performance of the KU15P when the VU7P is at 80\,\textcelsius{}. Magenta, cyan and grey dots on the right demonstrate the consistent thermal performance of three VU7Ps on three boards.}
\end{figure}

To avoid large temperature gradients, efforts are made to spread out the LUT oscillators evenly across the FPGA. The same number of LUT oscillators in each of the spread-out clusters are always enabled simultaneously during the test. The thermal gradients across the two SLRs of the VU7P are eliminated with the improved firmware design (Fig.~\ref{fig:finalperf}). 

At fan level 10, with the requirement that the FPGAs stay below 80\,\textcelsius{}, the revision~1 blade can consume 256\,W in total, with VU7P taking 140\,W, KU15P taking 86\,W and the SM and the rest of the CM components taking 30\,W. Figure~\ref{fig:finalperf} shows the thermal behavior of both FPGAs. Due to the size of the heatsinks and the FPGA package, the KU15P has worse thermal performance than the VU7P. Consistent thermal performance is observed on three CM boards. 

Because of the connection between the heatsinks, the KU15P is thermally impacted by the VU7P. When the temperature on the VU7P increases from 25\,\textcelsius{} to 80\,\textcelsius{}, the KU15P becomes 10\,\textcelsius{} warmer. To get to 80\,\textcelsius{}, the KU15P needs to consume 102\,W when the VU7P is at 25\,\textcelsius{}. It only needs to consume 86\,W when the VU7P is at 80\,\textcelsius{}. The VU7P, on the other hand, is not thermally impacted by the KU15P. 

The results are similar to the power estimation conducted during the design phase of the revision~1 blade. Power loss due to large currents running across the CM plane is measured to be about 12\,W at the maximum power.
 

\vspace{-7pt}

\section{Revision 2 and Future Plans}
\begin{figure}[htbp]
\begin{minipage}{0.50\textwidth}
\centering
\includegraphics[width=1.\textwidth]{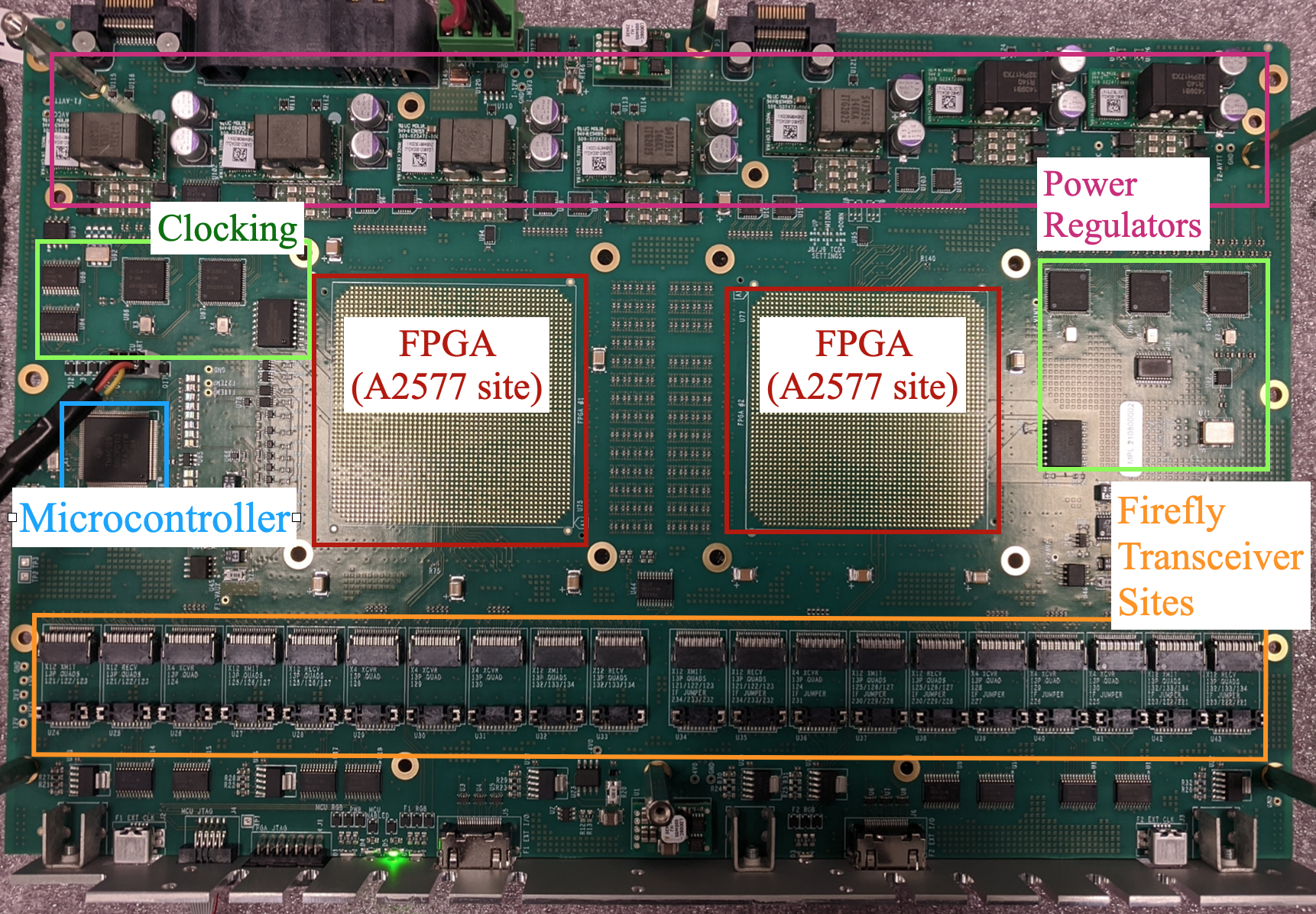}
\caption{\label{fig:rev2} The Apollo revision 2 CM without any FPGAs. Major components are labeled and highlighted in boxes.}
\end{minipage}
\qquad
\begin{minipage}{0.43\textwidth}
\centering
\includegraphics[width=1.\textwidth]{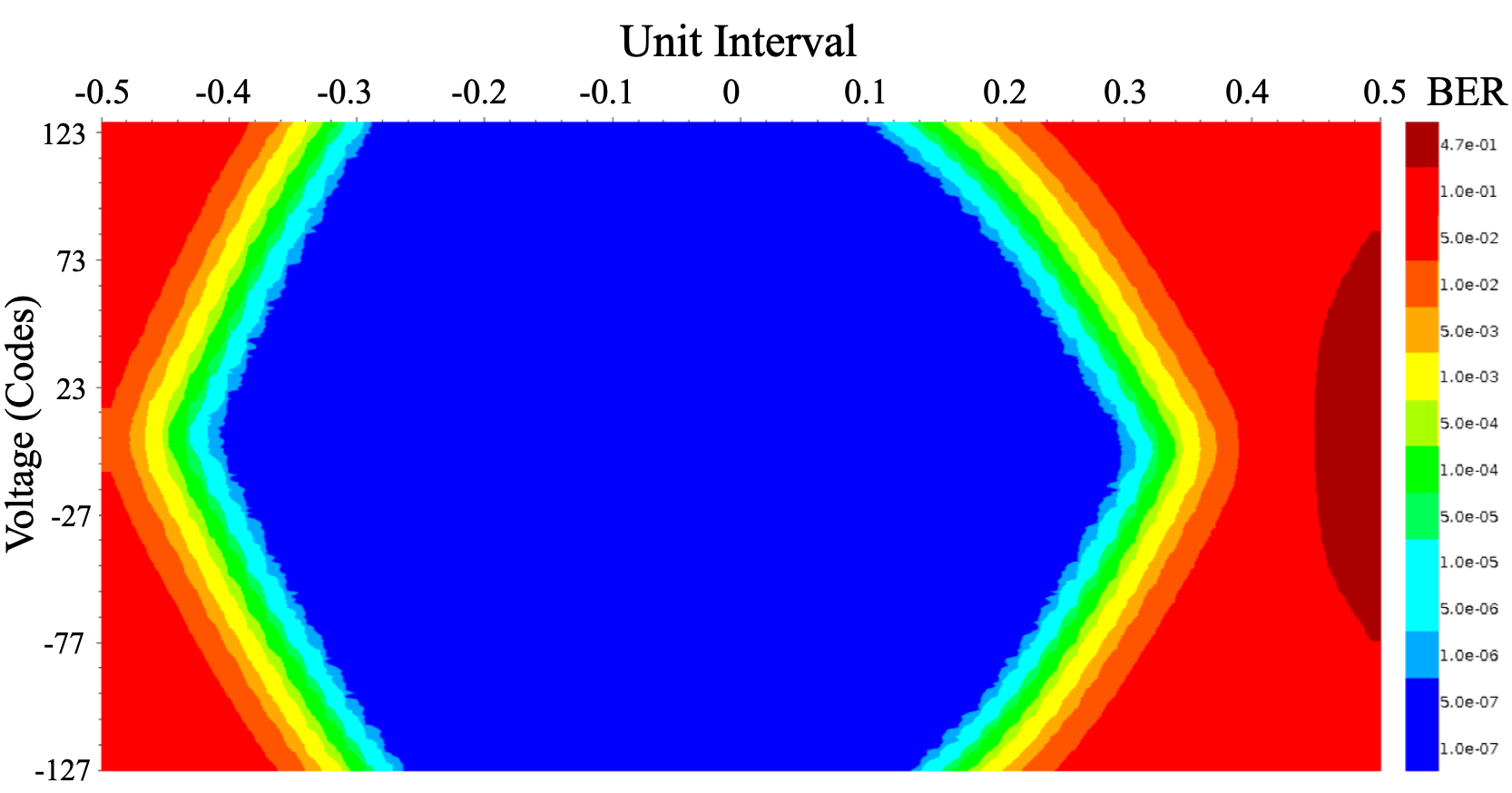}
\caption{\label{fig:rev2PRBS}A typical eye pattern with a large opening at a data rate of 25\,Gb/s with CDR on the revision~2 CM.}
\end{minipage}
\end{figure}

The revision~2 CM design (Fig.~\ref{fig:rev2}) features updates to clock signal distributions, more powerful FPGAs (two A2577 sites), power and thermal improvements, more adaptations for the optical links, and halogen-free material (EM-890). It has three \SI{70}{\micro\metre}-thick copper planes to reduce power loss in the distribution (revision~1 has two thinner planes). Sites for 14 Firefly optical engines (six 4-channel parts and eight 12-channel parts) are provided with up to 104 optical links with speed up to 25\,Gb/s. There are more links between the two FPGAs to provide the algorithms with options. The temperature sense diode outside the system monitors in each Xilinx FPGA is read out because of the temperature gradients observed in the revision~1 thermal tests.

The revision~2 CM is currently under evaluation. Preliminary studies show that a bit error rate less than $10^{-16}$ is achieved at 25\,Gb/s with large open eye diagrams (Fig.~\ref{fig:rev2PRBS}). Since the FPGA heatsinks are 3$\%$ larger than the VU7P heatsink in the revision~1 design and the FPGA package is larger, we anticipate each FPGA to be able to handle roughly 125--150\,W. For the CMs with a single FPGA, a larger heatsink up to twice the current size can be designed. 

Revision~2 of the SM is in production. Major design updates include an upgrade to Ultrascale+ Zynq SoC, a switch to halogen-free material (TerraGreen), and changes to the clock signal distribution.

\section{Summary}
The Apollo ATCA blade will be used for the IT-DTC and the TF in the CMS experiment at the HL-LHC. Two revisions of the designs are planned before the final production. The revision~1 design of the Apollo blade has demonstrated great link integrity and thermal performance. The evaluation tests for the revision~2 design are ongoing. Preliminary link integrity results are shown.


\acknowledgments
We would like to thank the authors of Ref.~\cite{heaterpaper} for the original VHDL code of the LUT oscillator used in our power tests. This work was supported by the National Science Foundation under grant no. NSF-PHY-1946735 and no. NSF-PHY-1912813. 



\end{document}